\documentstyle[12pt]{article}
\input epsf
\title{First quantum corrections for a hydrodynamics of a nonideal Bose gas}

\author{K.N.Ilinski$^{1,2}$\thanks{E-mail: kni@th.ph.bham.ac.uk}\ 
and\ A.S.Stepanenko$^{1,3}$\thanks{E-mail: ass@th.ph.bham.ac.uk}
\\
{\small\it $^{1}$ School of Physics and Space Research, University of
Birmingham,}
\\
{\small\it Birmingham B15 2TT, United Kingdom.} 
\\
{\small\it $^{2}$ Institute of Spectroscopy, Russian Academy of Sciences,} 
\\
{\small\it Troitsk, Moscow region, 142092, Russian Federation.}
\\
{\small\it $^{3}$ Theoretical Department, St-Petersburg Nuclear Physics Institute,}
\\
{\small\it Gatchina, St-Petersburg, 188350, Russian Federation.}
}
\date{ }

\begin{document}

\maketitle
\thispagestyle{empty}
\vskip -9cm
\vskip 9cm

\begin{abstract}
In the paper we consider a hydrodynamical description of a nonideal
Bose gas in one-loop approximation. We calculate an effective action which consists of mean field contributions and 
{\it first quantum correction}. 
This provides the equations of motion for the density and velocity of the gas where both mean field contributions and fluctuations are presented.
To fulfill the  calculation we make use of the formalism of functional 
integrals to map the problem to a problem of
quantum gravity to benefit from a well-developed technique in this field.
This effective action provides all correlation functions for 
the system and is a basis for a consideration of dynamics of the gas.  
Response functions are briefly discussed. Applications to the trapped bosons
are reviewed. Together with Ref.\cite{IS} the paper provide complete 
description of the condensate fraction and the deplition in the case of 
Bose condensed gases.
\end{abstract}

\section{Introduction}

Recently a lot of attention was attacted to the theory of a nonuniform Bose gas
and its collective excitations. The interest was stimulated by the success of the experimental obseravation of Bose-Einstein condensation systems of spin
polarized magnetically trapped atomic gases at ultra-low temperatures 
\cite{Exp1,Exp2,Exp3} and experimental studies of their collective 
properties~\cite{Exp4}. On the other hand, new exciting problems 
such as a  description of
the evolution of Bose condensate from relaxed trap, dynamics of 
a collapse of the condensate for Li$^7$ atoms, heating-cooling 
phenomena, various coherence effects for the 
condensate and so on are being posed both theoretically and experimentally.

Theoretically the evolution of the condensate and all related phenomena were
investigated in number of papers \cite{HC,RHB,EDC,Kagan1} but all 
these considerations were
formulated in Hartree-Fock (or mean field) approximation. It means that
the calculations were based on Nonlinear Schrodinger equation (NSE) or
Ginzburg-Gross-Pitaevskii equation (GPE) \cite{GP}. It is obvious that the GPE describes strongly interacting Bose condensed gas well since it is possible to ignore non-condensed fraction. However, as it was noted in \cite{IS}, this approximation fails to describe heating-cooling effects, fast dynamics
or fast external potential variations, fluctuation effects around the edges of a
trap and so on. To include the phenomena in the self-consistent
consideration quantum corrections to the GPE should be calculated.

Let us note that in the previous papers GPE plays double role.
First of all it describes the Bose condensate fraction. But it also
gives the dynamics of full density since in this approximation the deplition
was ignored. 
Based on GPE collective excitations and the deplition were considered in
Refs\cite{F1,F2,Stringari,Burnett1,WG} but the it was not self-consistent
and the interaction of the condensate fraction with the noncondensed one
have not been taken into account \cite{note5}. As a result,
to leave the mean field approximation the GPE equation can be generalized in two different ways --
we can write an equation for the condensate fraction as it was done in \cite{IS}
and its analogue for the full gas density.
In the paper \cite{IS} first (one-loop) quantum correction in the hydrodynamical
approximation was calculated to generalize the GPE for the
condensate fraction including effects of the interaction of the condensate
with collective excitations (noncondensed fraction). To calculate the deplition
in the framework we have to find Green functions for the corresponding
effective action. Being principally straightforward this is not an easy
technical problem. However there is a simpler way to find the diplition
as a difference between full density and the condensate density if the first is calculated in the same approximation (it is not difficult to see that the mean field equation for the density is once again GPE as it was noted above). 
In the paper we solve the problem and  give dynamical equations for the full density of the particles which generalizes GPE equation with an account of first quantum correction to the mean field picture. This allows us to calculate the deplition as well.

The paper is organized as follows.  In the next section
 we introduce functional integration to calculate the effective
hydrodynamic action for the system and define one-loop quantum corrections which
are responsible for the fluctuation effects.
These quantum corrections are evaluated using $\zeta$-function regularization
for functional determinants and an explicit and very simple 
expression for the
one-loop quantum corrections is presented. This permits in Section 3 the derivation of the 
quantum corrections to the equations of motion for the gas in closed and
explicit form readily used in numerical calculations. Here corrections 
to the Nonlinear Schr\"odinger equation (due to
the interaction of the mean field background with the fluctuations)
are obtained. Section 4 is devoted to a consideration of Green functions and response functions.
It is shown that all Green functions in the 1-loop approximation can be obtained from the effective action of Section 3 by functional differentiation. We conclude the paper with a discussion of the possible applications of our results. The analysis does not depend on the details of the
confining potential and can be used in a variety of problems. The only simplification  is common hydrodynamic approximation.

\section{Hydrodynamical picture}

As it was said above we consider nonideal Bose gas with
point-like interaction in external potential.
The Hamiltonian  of the system then can be written as
\begin{equation}
H = \frac{\hbar^{2}}{2 m} \int\! {\rm d}V\ \left[
\nabla \psi^{+} \nabla \psi +
\upsilon (x,t)\psi^{+}\psi +
4\pi l \psi^{+}\psi^{+}\psi\psi \right],
\label{H}
\end{equation}
where $\upsilon (x,t)$ is an external potential. For the case of experimentally trapped atoms the potential is supposed to be harmonic and can be varied or even switched off at some moment, say $t=0$. More precisely it means that for the case of magnetically trapped atoms 
$\upsilon =\theta (-t) a_{\perp}^{-4}(t)(\rho^{2}+\lambda^2 (t) z^2)$, $\rho^2=x^2+y^2$, ($a_\perp$ is oscillator lengths) has to be put in Eqn(\ref{H}).
Another parameter in Eqn(\ref{H}) is $l$ --- the s-wave scattering length  
for atoms in the system \cite{Osc}.
The operators $\psi^{+},\psi$ are Bose creation and annihilation operators.
This Hamiltonian is a basis for the calculation of quantum mechanical and thermodynamical quantities. For example,
the corresponding vacuum-vacuum transition amplitude is given by 
the following functional integral:
\begin{equation}
Z(\mu) =
\int\! {\rm D}\psi^{+} (x,\tau ){\rm D}\psi (x,\tau )\ \exp[iS/\hbar] \ ,
\label{Z}
\end{equation}
with the action
\begin{equation}
S = \int_{-\infty}^{\infty} {\rm d}\tau \int\! {\rm d}V
\left\{ i\hbar\psi^{+}\frac{\partial \psi}{\partial \tau}
- \frac{\hbar^{2}}{2 m}\left[\nabla \psi^{+} \nabla \psi +
(\upsilon - \frac{2 m}{\hbar^{2}}\mu)\psi^{+}\psi + 
4\pi l \psi^{+}\psi^{+}\psi\psi \right] \right\} \ .
\label{S0}
\end{equation}
where  the parameter $\mu$ is the chemical potential, controlling the number 
of particles in the system.

Now we will derive the hydrodynamic description directly from the functional integral formalism, which simplifies the description. We will consider zero temperature case but the generalisation for the case of finite temperatures
is straightforward (though cumbersome) using Keldysh technique \cite{Keldysh}.
As it is explained in Appendix A, we are 
looking for the effective action which describes all physical quantities for the system. For example, the effective action provides all
Green's functions in the same approximation used to calculate the effective action itself. We will obtain it in a hydrodynamic
one-loop approximation. To clarify the
description hydrodynamical variables, density and velocity, should be used.
More precisely, we start with the action (\ref{S0})
\begin{equation}
\frac{S}{\hbar} = \frac{1}{\hbar}\int_{-\infty}^{\infty} 
{\rm d}\tau \int\! {\rm d}V
\left\{i\hbar \psi^{+} \frac{\partial \psi}{\partial \tau}
- \frac{\hbar^{2}}{2 m}
\left[\nabla \psi^{+} \nabla \psi +
(\upsilon - \frac{2 m}{\hbar^{2}}\mu)\psi^{+}\psi + 
4\pi l (\psi^{+}\psi)^2 \right] \right\} \ ,
\end{equation}
and rescale variables as $\tau \rightarrow \frac{2m}{\hbar}\tau$, 
$\mu \rightarrow \frac{\hbar^2}{2m} \mu$, 
$l \rightarrow \frac{l}{4\pi}$
to get the following action form:
\begin{equation}
\frac{S}{\hbar} = \int_{-\infty}^{\infty} {\rm d}\tau \int\! {\rm d}V
\left\{ i\psi^{+} \frac{\partial \psi}{\partial \tau}
- \left[\nabla \psi^{+} \nabla \psi +
(\upsilon  - \mu)\psi^{+}\psi + 
 l (\psi^{+}\psi)^2 \right] \right\} \ .
\label{S1}
\end{equation}
(We will measure the action in terms of $\hbar$ everywhere below).
Now we change field variables to the hydrodynamic ones:
$$
\psi (x,\tau) = \sqrt{\rho(x,\tau)} e^{-i\varphi(x,\tau)}
\qquad , \qquad 
\psi^+ (x,\tau) = \sqrt{\rho(x,\tau)} e^{i\varphi(x,\tau)} \ .
$$
Then the action (\ref{S1}) takes the form (up to a complete derivative term):
\begin{equation}
S = \int_{-\infty}^{\infty} {\rm d}\tau \int\! {\rm d}V
\left\{ 
\frac{\partial \varphi}{\partial \tau}\rho -
(\nabla \sqrt{\rho})^2 - \rho (\nabla \varphi)^2 -
(\upsilon - \mu)\rho -
 l \rho^2 \right\} \ .
\label{S2}
\end{equation}
It is not difficult to see that the classical equations of motion for the
action lead to the equations:
\begin{equation}
\frac{\partial \varphi}{\partial \tau} -
 (\nabla \varphi)^2
-v  + \mu -
 2 l \rho + \frac{1}{\sqrt{\rho}}\nabla^2 \sqrt{\rho}  = 0 \ ,
\label{eqn1}
\end{equation}
\begin{equation}
- \frac{\partial \rho}{\partial \tau} +
2 \nabla (\nabla\varphi  \cdot\rho)
  = 0 \ ,
\label{eq2}
\end{equation}
which are equivalent to the Gross-Pitaevskii equation \cite{Stringari}
after the introduction of the velocity variable ${\bf c}=-2\nabla \varphi$
instead of $\varphi$. In velocity-density variables these equations look
as hydrodynamic equations for an irrotational compressible fluid:
\begin{equation}
\frac{\partial {\bf c}}{\partial \tau} + \nabla \left(
\frac{{\bf c}^2}{2}
+2v  - 2\mu +
4 l \rho - \frac{2}{\sqrt{\rho}}\nabla^2 \sqrt{\rho}\right)  = 0 \ ,
\label{eq3}
\end{equation}
\begin{equation}
\frac{\partial \rho}{\partial \tau} +
 \nabla ({\bf c}  \cdot\rho)
  = 0 \ .
\label{eq4}
\end{equation}

Now let us shift our variables by the zero-order mean field solution
(see Appendix A for details):
$$
\rho \rightarrow \rho + \sigma \qquad , \qquad 
\varphi \rightarrow \varphi + \alpha
$$
and we will keep up only terms including the square of $\sigma$ and $\alpha$ terms. Then the action  transforms to $S(\rho,\varphi) + S_1(\rho,\varphi,\sigma,\alpha)$
where the action $S_1$ has the form:
\begin{eqnarray}
S_1 &=& \int_{-\infty}^{\infty} {\rm d}\tau \int\! {\rm d}V\
\Biggl\{ 
\frac{\partial \alpha}{\partial \tau}\sigma -   
\rho (\nabla \alpha)^2 
-2\sigma \nabla \cdot \varphi \nabla \alpha
 -l \sigma ^2 
\nonumber\\
&&\qquad - \frac{1}{4\rho}(\nabla \sigma )^2  
 - \frac{1}{8\rho}
\left[
\frac{\Delta\rho}{\rho}
 - \frac{3(\nabla\rho)^2}{2\rho^2}
\right]\sigma^2 
\Biggr\} \ .
\label{S3}
\end{eqnarray}
Last two terms  vanish in the hydrodynamical limit since
$\rho$ is a large variable. Hence the action $S_1$ takes form:
\begin{equation}
S_1 = \int_{-\infty}^{\infty} {\rm d}\tau \int\! {\rm d}V
\left\{ 
\frac{\partial \alpha}{\partial \tau}\sigma -
\rho (\nabla \alpha)^2 
 -l \sigma ^2 -
2\sigma \nabla\varphi \cdot \nabla \alpha  
\right\} \ .
\label{S4}
\end{equation}
 Integrating in functional integral over the  field $\sigma$ we get  the action for 
$\alpha$ field only:
\begin{equation}
S_1 = \int_{-\infty}^{\infty} {\rm d}\tau \int\! {\rm d}V
\left\{ -\rho (\nabla \alpha)^2 +
\frac{1}{4l}\Bigl(\frac{\partial\alpha}{\partial \tau} 
- 2 \nabla \varphi \nabla \alpha\Bigr)^2 
\right\} \ .
\label{S5}
\end{equation}
The corresponding determinant is an inessential constant.

Now we will use the large  parameter $\rho$ (for harmonic trap 
$\rho\sim N^{2/5}l^{-3/5}$ if co-ordinates and $l$ are measured in 
harmonic length units $a_\perp$).
Let us introduce the following large number 
$\rho_0 \equiv {\rm max}\, \{\rho \}$ such that
$\tilde{\rho} = \frac{\rho}{\rho_0} \sim 1$, 
$t \equiv \tau \sqrt{4 l \rho_{0}}$, 
$\tilde{\bf v} = \frac{\nabla \varphi}{\sqrt{l \rho_{0}}} \sim 1$.
Then

\begin{eqnarray}
S_1\!\!& =&\!\! \sqrt{\frac{\rho_0}{4l}} 
\int_{-\infty}^{\infty} {\rm d}t \int\! {\rm d}V
\left\{ -\tilde{\rho} (\nabla \alpha)^2 +
\Bigl(\frac{\partial\alpha}{\partial t}
 - \tilde{\bf v} \nabla \alpha\Bigr)^2 
\right\} 
\nonumber\\
\!\!& \equiv&\!\! -\sqrt{\frac{\rho_0}{4l}} 
\int_{-\infty}^{\infty} {\rm d}t \int\! {\rm d}V\
A^{\mu \nu} \partial_{\mu}\alpha \partial_{\nu}\alpha.
\label{S6}
\end{eqnarray}
where the matrix $A$ has a form
$$
A= \left(  \begin{array}{cccc}
-1  &  \tilde{v}_1  &  \tilde{v}_2  &  \tilde{v}_3  \\
\tilde{v}_1 & \tilde{\rho}-\tilde{v}_1^2 & -\tilde{v}_1\tilde{v}_2  
&  -\tilde{v}_1\tilde{v}_3 \\
\tilde{v}_2 & -\tilde{v}_1\tilde{v}_2 &  \tilde{\rho}-\tilde{v}_2^2 
&  -\tilde{v}_2\tilde{v}_3  \\
\tilde{v}_3 & -\tilde{v}_1\tilde{v}_3 & -\tilde{v}_2\tilde{v}_3 
&  \tilde{\rho}-\tilde{v}_3^2  \\
\end{array} \right)
$$
with its determinant ${\rm det}\, A = -\tilde{\rho}^{3}$.

Our next step is to cast the action in the covariant form. To do this we introduce
auxiliary metric $\tilde{g}_{\mu\nu}$ such that 
$$
A^{\mu \nu} = \frac{\tilde{g}^{\mu\nu}}{\sqrt{-{\rm det}\, 
(\| \tilde{g}^{\mu\nu} \|)}} \ .
$$
One can easy to find the covariant metric from the equation above:
$$
\tilde{g}^{\mu \nu} = \frac{A^{\mu\nu}}{\sqrt{-{\rm det}\, (\| A^{\mu\nu} \|)}}
$$
that leads to the form:
$$
\|\tilde{g}^{\mu\nu}\|= \tilde{\rho}^{-3/2}
\left(  
\begin{array}{cccc}
-1  &  \tilde{v}_1  &  \tilde{v}_2  &  \tilde{v}_3  \\
\tilde{v}_1 & \tilde{\rho}-\tilde{v}_1^2 & -\tilde{v}_1\tilde{v}_2  
&  -\tilde{v}_1\tilde{v}_3 \\
\tilde{v}_2 & -\tilde{v}_1\tilde{v}_2 &  \tilde{\rho}-\tilde{v}_2^2 
&  -\tilde{v}_2\tilde{v}_3  \\
\tilde{v}_3 & -\tilde{v}_1\tilde{v}_3 & -\tilde{v}_2\tilde{v}_3 
&  \tilde{\rho}-\tilde{v}_3^2  \\
\end{array} \right) \ .
$$
Finally we get the metric $\|\tilde{g}_{\mu\nu}\|$:
$$
\|\tilde{g}_{\mu\nu}\|= \tilde{\rho}^{1/2}
\left(  
\begin{array}{cccc}
-\tilde{\rho} + \tilde{v}^2 &  \tilde{v}_1  &  \tilde{v}_2  &  \tilde{v}_3  \\
\tilde{v}_1 & 1 & 0  &  0 \\
\tilde{v}_2 & 0 &  1 &  0  \\
\tilde{v}_3 & 0 & 0 &  1  \\
\end{array} \right) \ .
$$
with the determinant 
$\tilde{g}\equiv {\rm det}\,(\|\tilde{g}_{\mu\nu}\|) = -\tilde{\rho}^{3}$.

In this metric the action takes a covariant form
\begin{equation}
S_1 = -\sqrt{\frac{\rho_0}{4l}} \int\! {\rm d}x\ \sqrt{-\tilde{g}}
 \alpha  \tilde{\mathop\Box^{ }} \alpha = \sqrt{\frac{\rho_0}{4l}}\int\! {\rm d}x\ 
\sqrt{-\tilde{g}} 
 \alpha  \frac{1}{\sqrt{-\tilde{g}}} \partial_{\mu} \tilde{g}^{\mu \nu}  
\sqrt{-\tilde{g}} 
\partial_{\nu} \alpha \ .
\label{S8}
\end{equation}
This is the central equation of the paper.

Now the effective action is 
$\Gamma _{1} = - {\rm tr}\, \ln [(\rho_0/4l)^{1/2} 
\tilde{\displaystyle\mathop\Box^{ }} ]/2$
with $\tilde{\displaystyle\mathop\Box^{ }} = -\frac{1}{\sqrt{-\tilde{g}}} \partial_{\mu} 
\tilde{g}^{\mu \nu}  \sqrt{-\tilde{g}} \partial_{\nu} $.
Let us again make use the existence of large parameter in the system.
Indeed, the multiplier $(\rho_0/4l)^{1/2}$ gives a possibility to
express main contributions to the determinant such as (see Appendix B):
$$
{\rm tr}\, \ln [(\rho_0/4l)^{1/2} 
\tilde{\displaystyle\mathop\Box^{ }} ] = {\rm tr}\, \ln [ 
\tilde{\displaystyle\mathop\Box^{ }} ] +
\frac{1}{2} {\rm tr}\, \ln (\rho_0/4l) 
\Bigl(\Phi_0 (\tilde{\displaystyle\mathop\Box^{ }} )
 - L(\tilde{\displaystyle\mathop\Box^{ }}) \Bigr)
$$
$$
\sim \frac{1}{2} {\rm tr}\, \ln (\rho_0/4l) 
\Bigl(\Phi_0 (\tilde{\displaystyle\mathop\Box^{ }} )
 - L(\tilde{\displaystyle\mathop\Box^{ }}) \Bigr)
$$
since in our regularization 
${\rm tr}\, \ln [ \tilde{\displaystyle\mathop\Box^{ }} ]$ is order of Seeley 
coefficient $\Phi_0(\tilde{\displaystyle\mathop\Box^{ }})\equiv
 \int\!  {\rm d}x{\rm d}\tau\ \sqrt{-g} \Psi_0$.
Returning to the initial variables all curvature tensors  and the metric
are written in the ``physical" variables (i.e. without tildes).
Moreover $\ln (\rho_0/4l) \gg \ln (\tilde{\rho})$ so that we can substitute
$\ln (\rho/4l)$ instead of $\ln (\rho_0/4l)$. Summarizing, 
we obtain an expression for the first quantum correction to the
effective action:
\begin{equation}
\Gamma _{1} = \Gamma _{1}^\prime + \Gamma _{1}^{\prime\prime} 
\label{Gamma1}
\end{equation}
 where 
$$
\Gamma _{1}^\prime = -\frac{1}{4} \int\!  {\rm d}x{\rm d}\tau\ \sqrt{-g}
\ln (\rho/4l) \Psi_0(\Box ) \ ,
$$
$$
\Gamma _{1}^{\prime\prime} = \frac{1}{4} \ln (\rho_0/4l) L(\Box ) \ .
$$
with the metric 
$$
\|g_{\mu\nu}\|= \rho^{1/2}
\left(  
\begin{array}{cccc}
-\rho + v^2 &  v_1  &  v_2  &  v_3  \\
v_1 & 1 & 0  &  0 \\
v_2 & 0 &  1 &  0  \\
v_3 & 0 & 0 &  1  \\
\end{array} \right) \ .
$$
and $v_i\equiv 2\partial_i\varphi$.

\section{Quantum corrections to equations of motion}
In this section we derive from (\ref{Gamma1}) required quantum corrections to equations of motion (see Appendix A).
To make the consideration self-consistent  we should not differentiate 
$ \ln (\rho/4l)$ because in our approximation it behaves like a constant.
In that case, the action we have to vary is covariant again.
we find:
$$
\frac{\delta \Gamma_{1}}{\delta \rho} = 
\frac{\delta g_{\mu\nu}}{\delta \rho}
\frac{\delta \Gamma_{1}^\prime}{\delta g_{\mu\nu}}
 +\frac{1}{4} \ln (\rho/4l) \frac{\delta L(\Box )}{\delta \rho} 
$$
$$
\frac{\delta \Gamma_{1}}{\delta \varphi} = 
\frac{\delta g_{\mu\nu}}{\delta \varphi}
\frac{\delta \Gamma_{1}^\prime}{\delta g_{\mu\nu}}
 +\frac{1}{4} \ln (\rho/4l) \frac{\delta L(\Box )}{\delta \varphi} \ .
$$
Now,
$$
\frac{\delta \Gamma_{1}^\prime}{\delta g_{\mu\nu}} =
-\frac{1}{4} \ln(\rho/4l) \frac{\delta }{\delta g_{\mu\nu}}
\int \!{\rm d}x\ \sqrt{-g} \Psi_{0} (\Box ) = 
-\frac{1}{4} \ln(\rho/4l) \frac{\delta }{\delta g_{\mu\nu}} \Phi_{0} (\Box )
$$
Let us remember that
$$
\Phi_{0} (\Box ) =
\frac{1}{(4\pi)^2} \int \!{\rm d}x\ \sqrt{-g}
\left( -\frac{1}{30}\nabla^2 R + \frac{1}{72}R^2 -
\frac{1}{180}R_{\mu\nu} R^{\mu\nu}
 + \frac{1}{180}R_{\mu\nu\sigma\rho} R^{\mu\nu\sigma\rho} \right) 
$$
We note that one of the terms above is irrelevant:
$$
\int \!{\rm d}x\ \sqrt{-g} \nabla^2 R = 
\int \!{\rm d}x\ \sqrt{-g} \frac{1}{\sqrt{-g}}
\partial_{\mu} \sqrt{-g} g^{\mu\nu} \partial_{\nu} R
 = \int \!{\rm d}x\  \partial_{\mu} (\sqrt{-g}\partial^{\mu} R)
$$
since this is just a complete derivative. Hence for our purposes it is sufficient to consider only 
$$
\Phi_{0} (\Box ) =
\frac{1}{(4\pi)^2} \int \!{\rm d}x\ \sqrt{-g}
\left(\frac{1}{72}R^2 -
\frac{1}{180}R_{\mu\nu} R^{\mu\nu}
 + \frac{1}{180}R_{\mu\nu\sigma\rho} R^{\mu\nu\sigma\rho} \right) \ .
$$

Long but quite straightforward calculations lead us to the following result for the
functional derivative (see Appendix C for details of the calculation):
\begin{eqnarray*}
\frac{\delta }{\delta g^{\mu\nu}} \Phi_{0} (\Box )\!\!& =\!\! &
 \frac{1}{36(4\pi)^2}
\sqrt{-g} 
\Biggl\{
 - \frac{1}{2} g_{\mu\nu} 
\Bigl[ 
   \frac{1}{2}R^2
 - \frac{1}{5} R_{\sigma\rho}R^{\sigma\rho} 
 + \frac{1}{5}R_{\sigma\rho\alpha\beta}R^{\sigma\rho\alpha\beta}
\Bigr] 
\\
\!\!& +\!\! & 
   R_{\mu\nu}R
 - \frac{2}{5}R_{\mu\sigma}R_{\nu}^{\,\cdot\, \sigma} 
 + \frac{2}{5}R_{\mu\sigma\rho\alpha}R_{\nu}^{\,\cdot\,\sigma\rho\alpha}
 - g_{\mu\nu} \Box R
 + \frac{1}{2} \{\nabla _{\mu},\nabla_{\nu}\} R  
\\
\!\!& +\!\! &
\frac{1}{5} 
\Bigl[ 
 - 2\nabla^{\sigma}\nabla_{\nu} R_{\mu\sigma}
 + \Box R_{\mu\nu}
 + g_{\mu\nu}\nabla^{\sigma}\nabla^{\rho} R_{\sigma\rho} 
\Bigr]
 + \frac{4}{5} 
   \nabla^{\rho}\nabla^{\sigma}R_{\mu\sigma\rho\nu} 
\Biggr\}
\end{eqnarray*}
taking into account
$$
\frac{\delta }{\delta g_{\mu\nu}} \Phi_{0} (\Box ) =
- g^{\mu\sigma}g^{\nu\rho}\frac{\delta }{\delta g^{\sigma\rho}} 
\Phi_{0} (\Box )  \ .
$$
This means that the equations of motions with 1-loop quantum correction 
take the form:
\begin{equation}
\frac{\partial \varphi}{\partial \tau} -
 (\nabla \varphi)^2
-v  + \mu -
 2 l \rho + \frac{1}{\sqrt{\rho}}\nabla^2 \sqrt{\rho} 
+\frac{\delta g_{\mu\nu}}{\delta \rho}
\frac{\delta \Gamma_{1}^\prime}{\delta g_{\mu\nu}}
 +\frac{1}{4} \ln (\rho/4l) \frac{\delta L(\Box )}{\delta \rho} = 0 \ ,
\label{eqncorr1}
\end{equation}
\begin{equation}
- \frac{\partial \rho}{\partial \tau} +
2 \nabla (\nabla\varphi  \cdot\rho)
+ \frac{\delta g_{\mu\nu}}{\delta \varphi}
\frac{\delta \Gamma_{1}^\prime}{\delta g_{\mu\nu}} 
 +\frac{1}{4} \ln (\rho/4l) \frac{\delta L(\Box )}{\delta \varphi} = 0 \ ,
\label{eqncorr2}
\end{equation}
These two equations (\ref{eqncorr1}) and (\ref{eqncorr2}) replace the mean field
hydrodynamic equations (\ref{eqn1},\ref{eq2}) for the density and velocity of the
Bose gas and are the main result of the paper. They contain contributions
of fluctuations to the dynamics of the gas. For example, we
 note that the additional quantum pressure is due to dependence of the
Bogoliubov particles determinant on the density of the condensate while
 local density changes  (see continuity equation (\ref{eqncorr2})) comes
from phase-dependence of the determinant. There is no difficulty to see that the latter correction obeys particle conservation law (that means it is a divergence)\cite{Volovik}.

To complete this section we consider  the stationary
situation when the equilibrium condensate velocity is equal to zero there is no
local particle transfer.  However we think that 
equations (\ref{eqncorr1}) and (\ref{eqncorr2}) are of the main
interest in non-equilibrium problems such as evolution of the gas under
changing of  trap shapes and so on. In equilibrium case of trapped atomic gases, the velocity field is equal to zero
as are all time derivatives in the equations of the previous section.
This allows us to greatly simplify the equations and find quantum corrections
in closed and compact form.
Indeed, it is cumbersome but straightforward to check that in the regime there is no
corrections to continuity equation (\ref{eq4}), $L(\Box)=1$, and there is only the contribution
to the quantum pressure term in equation (\ref{eq3}).
This means that the equation of motion take the form:
\begin{equation}
\frac{{\bf c}^2}{4}
+\upsilon  - \mu +
2 l \rho
 - \frac{1}{\sqrt{\rho}}\nabla^2 \sqrt{\rho}
 + \delta P  = 0 \ ,
\label{NS}
\end{equation}
where the correction to quantum pressure $\delta P$ has the form:
\begin{eqnarray}
\delta P = 
\frac{\ln (\rho/4l)}{48(4\pi)^2 \sqrt{\rho}} 
\Biggl( 
& &\!\!\!\!
 \ \ \frac{43}{20} \Delta ^2 \ln \rho  
 + \frac{43}{20} (\partial_i \ln \rho)(\partial_i \Delta \ln \rho) 
\nonumber\\
  & &\!\!\!\!
 + \frac{27}{10} 
   (\partial_i \partial_k \ln \rho)(\partial_i \partial_k \ln \rho)
 - \frac{87}{80} (\Delta \ln \rho)^2
\nonumber\\
& &\!\!\!\!
 - \frac{67}{80} (\partial_i \ln \rho)(\partial_k \ln \rho)
   (\partial_i \partial_k \ln \rho)  
\nonumber\\ 
  & &\!\!\!\! 
 - \frac{197}{160} (\partial_i \ln \rho)^2 \Delta \ln \rho
 - \frac{417}{1280} [(\partial_i \ln \rho)^2]^2 
\Biggr)
\label{delta}
\end{eqnarray}
where roman indices correspond to spatial derivatives.

This implies the mean field approximation is applicable at the center of 
the trap where the correction $\delta P$ is negligible.
On Fig.1 we plot the contribution in the central region of the
trap. 
Since the contribution is particularly important near to the classical
turning points we do not 
make use the Thomas-Fermi approximation and exploit more
correct analytic expression. 
The
profile of trapped condensate was obtained in Ref.\cite{IM}  in the mean field approximation as an interpolation
between small density and high density expansions; this demonstrates very good agreement with numerical solution of the corresponding mean field equation \cite{DS}. The expression for the profile with $N$ particles has the form:
\begin{equation}
\rho (r) = \frac{1}{2 l}W\left[2l N \exp(4C-r^2)\right],
\label{fe}
\end{equation}
where $W(x)$, defined as the principal branch (regular at the origin)
solution to the Eq. $We^W=x$ \cite{FSZ},
is  a standard MAPLE function \cite{MP};
Constant $C$ in  Eqn.(\ref{fe}) is to be determined from the normalization
condition $\int \,{\rm d}^3x\ \rho (r)= N$ \cite{note}. The profile of the condensate density calculated from eq.(\ref{fe}) is shown in Fig.1 for
$N=10^5$ atoms.

Two points should be noted. Firstly, the magnitude of the
correction in the central region of the trap is of order of $10^{-6}$ relative to the main interaction term (see Figs 2) (the contribution in 
the intermediate region is shown on Fig.3). Secondly, the sign
of the correction term changes as can be seen from Fig.4. This means that 
the character of exchange interaction due to excited states depends on the density
of the condensate and leads to an effective attraction for density higher than
some value and to an effective repulsion for a smaller density.
However we have to remind that the leading contributions in the region are much larger than the corrections and after a self-consistent solution of the corrected Nonlinear Schr\"odinger equation is performed the profile of the 
full density will not change significantly in this region of the trap.

Since the sign of the contribution $\delta P$ became positive, 
the value of this term starts to increase and becomes quite considerable.
However, at some point, our hydrodynamic approximation has to fail
and we need other correction terms which will  stop the increase. Although we cannot
treat them properly it is possible to give an approximation for this region
based  on the low density expansion. Up to now we dealt with a 
high density expansion denoted $\delta P_{h}$. 
It is easy to calculate the leading contribution
in low density expansion that controls the behavior of depletion
 interaction
correction at large distances. In section 2 we saw that this correction corresponds to sum of all the one-point diagrams for the system with 
the action (\ref{S0}).
In the low density regime the main contribution $\delta P_{l}$ is due to a one 
loop diagram. It means that the correction to eqn. (\ref{NS})
$\delta P_{l}$ is 
\begin{equation}
\delta P_{l} = 4 l \delta \rho \ .
\label{pl}
\end{equation}
It is hard to give exact expression for the deplition in this region since the
comprehensive solution is required. We estimate the depletion as one fourth
of the local density of the particles and use this estimate in Fig.5.
It  shows plots of  $\delta P_{l}$ and $\delta P_{h}$ which let us estimate
the contribution in the intermediate density region. It is easy to see that the two
curves meet at point $r\sim 6.13$ where the density of the condensate 
$\rho \sim 1$ and both (high and low density) approximations fails. 
This however let us estimate the magnitude of the correction.
It is obvious that the crack at the figure should be smoothed
and this comes from high order corrections. We cannot calculate them explicitly
and the only interpolation should be used.
The magnitude of the correction to the Nonlinear Schr\"odinger equation is
 $\sim 1$ and in the region is compared with both kinetic and repulsion energies as it is shown in Fig.6. 

We see that the only region where the corrections are not small is the
boundary region. Here the corrections are order of both kinetic and interaction energies
and can play significant role in the determination  of the profile of the condensate
close to classical turning points. The Thomas -Fermi approximation does not work in the region and hence both Stringari \cite{Stringari} and Wu and Griffin \cite{WG} approximations are inadequate. This explain the discrepancy of our results. 

Other point to note is that the form of Eqn.(\ref{NS}) coincides with
the form of the corresponding equation in the case of the condensate of
the gas \cite{IS}, i.e. there is only one generalization of GP equation 
for the stationary case. However, solutions for the condensate density 
and full density can be different since  the generalized GP equations 
is highly nonlinear and its solution crucually depends on initial conditions.

\section{Green functions, response functions and the like}

Equations of motion and quantum corrections for them are the central questions of previous consideration.  However in many applications related to experiment
it is very important to analyze  various response functions and form-factors.
They can be expressed in terms of  some combinations of Green functions of the theory.
That is why in this section we shortly consider a calculation of Green functions in the effective action formalism. 

As it is shown in Appendix A the effective action (in any self-consistent approximation) allows the evaluation of {\it all} Green functions in the same
approximation by just taking of variational derivatives. Since the effective action
in 1-loop approximation was obtained the problem of Green function calculation
is the problem of functional differentiation only \cite{note3}.

We now formalize all said above and give formulas to evaluate Green functions
in effective action formalism.
Let $W(J)$ being generating functional of connected Green functions. Then 
quantities
\begin{equation}
        W_n(x_1,\dots,x_n) = \frac{\delta}{\delta J(x_1)}\dots
        \frac{\delta}{\delta J(x_n)}\,W(J)
\label{G1}
\end{equation}
are connected Green functions in the external field $J$. The effective action 
is defined then by the Legendre transformation of $W$:
\begin{equation}
        \Gamma(\alpha) = W(J(\alpha)) - \alpha J(\alpha)\ ,\qquad
        \alpha = \frac{\delta W(J)}{\delta J}
\label{G2}
\end{equation}
where the function
$\alpha(x) = \langle\hat\varphi(x)\rangle = W_1(x;J)$ is the first 
connected Green function (\ref{G1}). It is easy
to see that the functions $\alpha$ and $A$ are related by the 
second relation (\ref{G2}). Then the quantities
\begin{equation}
        \Gamma_n(x_1,\dots,x_n) = \frac{\delta}{\delta \alpha(x_1)}\dots
        \frac{\delta}{\delta \alpha(x_n)}\,\Gamma(\alpha)
\label{G3}
\end{equation}
define 1PI (one-particle-irreducible) Green functions for the theory with
mean field $\alpha(x) = \langle\hat\varphi(x)\rangle$. Knowledge of 1PI Green function is equivalent to knowledge of any (whole or connected) Green functions for the
corresponding system. For the first 1PI 
Green function we have from (\ref{G2})
\begin{equation}
        \Gamma_1(x) = - J(x) \ .
\label{G4}
\end{equation}
All connected Green functions can be expressed in terms of 1PI ones. 
Indeed, differentiating (\ref{G4}) $J$ we obtain
\begin{equation}
        W_2\Gamma_2 = - 1\ ,\qquad W_2 = - \Gamma_2^{-1}
\label{G5}
\end{equation}
or, in expanded form,
\begin{equation}
        \int\!{\rm d}z\ W_2(x,z)\Gamma_2(z,y) = - \delta(x-y) \ .
\label{IP}
\end{equation}
Differentiating now the second relation in (\ref{G5}) on $J$ and using the 
following rule
$$
        \frac{\delta}{\delta J} = - \Gamma_2^{-1} \frac{\delta}{\delta \alpha}
$$
we can derive expressions for all the higher connected Green functions through
the 1PI ones. For example, for the third connected function we have
$$
        W_3 = - \left[\Gamma_2^{-1}\right]^{3}\Gamma_3
$$
or
\begin{equation}
        W_3(x_1,x_2,x_3) = - \int\!{\rm d}y_1{\rm d}y_2{\rm d}y_3\ 
        \Gamma_2^{-1}(x_1,y_1)\Gamma_2^{-1}(x_2,y_2)
        \Gamma_2^{-1}(x_3,y_3)\Gamma_3(y_1,y_2,y_3)
\label{IP2}
\end{equation}
and so on. It means that since we know 1PI functions we have to solve
the differential equation (\ref{IP}) for the two-point correlation function
and then find all other connected Green functions by integration of 1PI functions with
two-point correlators as in Eqn.(\ref{IP2}).

Putting $J=0$ in (\ref{G4}) we get equations of motion for $\rho$ and 
$\varphi$ (\ref{eqncorr1},\ref{eqncorr2}). To obtain, for example, the Green function 
$\langle(\rho(x)-\langle\rho (x)\rangle )(\rho(y) - \langle \rho (y)\rangle ) \rangle$ we twice differentiate the effective
action on $\rho$, substitute solution of equations of motion and finally 
invert the result in the sense of the kernel of a integral operator.

As an example of usage of this technique we calculate two-point Green function in the mean field approximation (tree or 0-loop approximation). 
In this approximation the effective action has the form:
\begin{equation}
\Gamma = \int_{-\infty}^{\infty} {\rm d}\tau \int\! {\rm d}V
\left\{ 
\frac{\partial \varphi}{\partial \tau}\rho -
(\nabla \sqrt{\rho})^2 - \rho (\nabla \varphi)^2 -
(\upsilon - \mu)\rho -
 l \rho^2 \right\} \ .
\label{ac0}
\end{equation}
such that, in the equilibrium hydrodynamic picture, the matrix of second variational derivatives
of the effective action can be written as
$$
\delta^2 \Gamma = 
\left(  
\begin{array}{cc}
\displaystyle\frac{\delta^2 \Gamma}{ \delta \rho (x) \delta \rho (y)}  &  
\displaystyle\frac{\delta^2 \Gamma}{ \delta \rho (x) \delta \varphi (y)}
 \\[3mm]
\displaystyle\frac{\delta^2 \Gamma}{  \delta \varphi (x)  \delta \rho (y)}  &  
\displaystyle\frac{\delta^2 \Gamma}{ \delta \varphi (x) \delta \varphi (y)} \\
\end{array} 
\right)
= 
\left(  
\begin{array}{cc}
-2 l  &  
-\displaystyle\frac{\partial }{ \partial t} \\
\displaystyle\frac{\partial }{ \partial t} &  
2 \nabla \cdot \rho \nabla \\
\end{array} 
\right)  \  .
$$
where $\rho\equiv\langle\hat\rho\rangle$. The
matrix of two-point connected correlation functions of fluctuations
$\hat{\tilde\rho} (x) = \hat\rho (x) - \langle \hat\rho (x) \rangle$ 
and 
$\hat{\tilde\varphi} (x)= \hat\varphi (x) - \langle \hat\varphi (x) \rangle$
$$
G = 
\left(  
\begin{array}{cc}
\langle \hat{\tilde\rho } (x) \hat{\tilde\rho } (y) \rangle  &  
\langle \hat{\tilde\rho } (x) \hat{\tilde\varphi } (y) \rangle \\[1.5mm]
\langle \hat{\tilde\varphi } (x) \hat{\tilde\rho } (y) \rangle  &  
\langle \hat{\tilde\varphi } (x) \hat{\tilde\varphi } (y) \rangle \\
\end{array} 
\right)
$$
is defined as a solution of the equation
$$
\delta^2\Gamma  \cdot  G  = - {\rm I} \ .
$$
Solving the equation we get the following expressions for the correlators
$$
\langle \hat{\tilde\rho } (x) \hat{\tilde\rho } (y) \rangle = 
\delta (x-y) - 
\frac{1}{2l}\frac{\partial }{\partial t} \langle \hat{\tilde\varphi } (x) 
\hat{\tilde\rho } (y) \rangle 
$$
$$
\langle \hat{\tilde\rho } (x) \hat{\tilde\varphi } (y) \rangle = 
-\frac{1}{2l}\frac{\partial }{\partial t}
\langle \hat{\tilde\varphi } (x) \hat{\tilde\varphi } (y) \rangle 
$$
and the equation for the phase-phase correlator 
$\langle \hat{\tilde\varphi } (x) \hat{\tilde\varphi } (y) \rangle$:
$$
\left( \frac{1}{2l}\frac{\partial^2 }{\partial t^2}
 - 2 \nabla \cdot \rho \nabla
\right) \langle \hat{\tilde\varphi } (x)  \hat{\tilde\varphi } (y) \rangle
 = \delta (x-y)
$$
(here $x$ denotes both space and time variables).
It means that the phase-phase correlator is the Green function 
of the wave equation.

In summary, various response functions, form-factors and Green functions
are evaluated from effective action in any given approximation taking
variational derivatives and integrating them with the two-point correlation function. This can be done in general  using formulas of this section.

\section{Conclusion}

In the paper a self-consistent approach to the calculation of
1-loop quantum corrections to the Gross-Pitaevskii equation for a Bose gas density due to fluctuations around mean field configuration are calculated. To do this the hydrodynamic approximation was used. In this approximation excitations are equivalent to
sound waves on a mean field  background. This opens the possibility to use {\it methods of quantum gravity and theory of quantum gauge fields} where the problem of calculation of effective action for gravitational and gauge backgrounds 
(or more precise, quantum corrections due to other quantum fields) is common problem. 

Many of the methods to approach the problem were developed during the last four decades. They are $\zeta$-function
regularization for determinants of operators, 
Schwinger-De-Witt-Seeley expansion of heat kernels \cite{Schwinger,DW,Seeley},
covariant methods of calculation of Seeley's coefficients and
the covariant perturbation technique for effective actions \cite{BV,BGVZ}.
Although the methods are well-known in field theory they are not so
familiar in condensed matter physics and the theory of coherent systems. 
An accurate account of corrections is very complicated even for a few first orders and  the solutions are obtained by using the covariant perturbation technique and curvature expansions. On the way effects nonlocal in space and time appeared. 
In quantum gravity framework such effects play significant role in the gravitational collapse problem, Hawking radiation and so on. 

However to benefit from it sometimes very general and 
only basic information about principle scales in systems in question is
required.
Indeed, this is the only information needed to extract leading logarithmic contributions while the calculation of other corrections is much more complicated. 
That is why we think that the developed approach can be widely used in many
problems which have not much to do with Bose-condensation of trapped atoms
or liquid Helium. 

At the end let us stop on other applications of the presented method.
It is easy to imagine other condensed matter examples of problems treatable by the same technique. Clearest and simplest of them are
antiferromagnets, Josephson arrays and superconductors in nonuniform (in space and time) external magnetic fields. However many of other physical systems
in a non-equilibrium background are potential field of applications.

Returning to Bose condensation, the paper together with our
previous work \cite{IS} where the generalization of Gross-Pitaevskii 
equation for the condensate fraction was given allows to consider the
condensate fraction and the deplition on the same basis since the deplition is
a difference of the full density and the condensate density. However, the results of this paper can be used even without refering to the Bose 
condensation phenomena.

\vspace{1cm}

\noindent
{\it Acknowledgment.}  
We are very grateful to Mike Gunn for frequent stimulating discussions and 
valuable comments.
We also want to thank Keith Burnnett, Dima Vasilevich, Martin Long, Ray Jones, 
Igor Lerner, Pavel Kornilovich and Ely Klepfish for the discussions of the problem. 
This work was supported by the Grant of Russian Fund of Fundamental
Investigations N 95-01-00548 and by the UK EPSRC Grants GR/L29156,
GR/K68356.

\vspace{1cm}

\noindent
{\huge \bf Appendices}

\appendix

\def\theequation{\Alph{section}.\arabic{equation}}
\setcounter{equation}{0}

\section{Effective action}

In this Appendix we state some basic information concerning the formalism of an effective action following \cite{Avr}. 
Let us consider a set of fields
$\varphi^i\equiv\varphi^A(x)=\{\psi(x),\psi^{+}(x)\}$.
 The classical dynamics of these
fields is described by the classical equations of motion
\begin{equation}
{\delta S \over \delta\varphi^A(x)}=0
\label{1}
\end{equation}
derived from the classical action functional
\begin{equation}
S(\varphi)=\int\!{\rm d}x\ 
{\cal L}\left(\varphi(x),\partial_\mu\varphi(x)\right) \ .
\label{2}
\end{equation}

After quantization the fields become field operators $\hat \varphi$ acting
on some state vectors of Fock space. Let us suppose
that there exist some initial $|{\rm in}\rangle$ and final $\langle {\rm out}|$ 
vacuum states  defined in some appropriate way. 
The vacuum-vacuum transition amplitude can be
presented as a Feynman path integral
\begin{equation}
\langle{\rm out}|{\rm in}\rangle\equiv Z(J)=\int\! {\rm D}\varphi\
\exp\left\{\frac{i}{\hbar}\left(S(\varphi)+J_i\varphi^i\right)\right\}
\label{3}
\end{equation}

Here we introduced the classical sources $J_i$ to investigate the linear
reaction of the system on the external perturbation. All the content of quantum
field theory with all quantum effects is contained in the following Green
functions
\begin{equation}
\left\langle\hat\varphi^{i_1}\cdots\hat\varphi^{i_n}\right\rangle\equiv
{\left\langle{\rm out}\right|{\rm T}\left(\hat\varphi^{i_1}\cdots\hat\varphi^{i_n}\right)
\left|{\rm in}\right\rangle
\over \left\langle{\rm out}|{\rm in}\right\rangle}
\label{4}
\end{equation}
where T means the chronological ordering operator, i.e. the fields must be
arranged from left to right in order of decreasing time arguments.

It is easy to show that all these Green functions can be obtained by the
functional differentiation of the functional $Z(J)$, that is called, therefore,
the generating functional. Moreover, factorizing the unconnected contributions
one comes to generating functional for connected Green functions 
$W(J)=-i\hbar\ln Z(J)$
\begin{equation}
\left\langle\hat\varphi^{i_1}\cdots\hat\varphi^{i_n}\right\rangle
=\left(\frac{\hbar}{i}\right)^{n}\exp(-\frac{i}{\hbar}W)
{\delta^n\over \delta J_{i_1}\cdots \delta J_{i_n}}\exp(\frac{i}{\hbar}W).
\label{5}
\end{equation}

The lowest connected Green functions have special names: the mean field
\begin{equation}
\left\langle\varphi^i\right\rangle\equiv\phi^i(J)={\delta W\over \delta J_i}
\label{6}
\end{equation}
and the propagator
\begin{equation}
\left\langle\left(\hat\varphi^i-\phi^i\right)\left(\hat\varphi^k-\phi^k\right)\right\rangle
\equiv -i\hbar G^{ik}(J)=-i\hbar {\delta^2 W\over \delta J_i \delta J_k}
\label{7}
\end{equation}

Further, the connected Green functions are expressed in terms of vertex
functions. The generating functional for vertex functions is defined then by
the Legendre functional transform
\begin{equation}
\Gamma({\bf \phi})=W(J({\bf \phi}))-J_k({\bf \phi}){\bf \phi}^k
\label{8}
\end{equation}
This is the most important object in quantum field theory. It contains all the
information about the quantized fields.
\begin{enumerate}
\item First, one can show that it satisfies the equation
\begin{equation}
\Gamma_{,i}({\bf \phi})\equiv {\delta \Gamma\over \delta {\bf \phi}^i}=-J_i({\bf \phi})
\label{9}
\end{equation}
and, therefore, gives the effective equations for the mean field. These
equations replace the classical equations of motion and describe the effective
dynamics of background fields taking into account all quantum corrections. Thus
$\Gamma({\bf \phi})$ is called usually the {\it effective action}.
\item Second, it determines the full or exact propagator of quantized
fields
\begin{equation}
-\Gamma_{,ik}G^{kn}=\delta^n_i
\label{10}
\end{equation}
where $\delta^n_i\equiv\delta^B_A\delta(x-y)$,
and the vertex functions
\begin{equation}
\Gamma_n\equiv\Gamma_{,i_1\cdots i_n}, \qquad (n\ge 3)
\label{11}
\end{equation}
This means that any $S$-matrix amplitude, or any Green function, is expressed
in terms of propagator and vertex functions that are determined by the
effective action.
\item At last, when the test sources vanish the effective action is just
the vacuum amplitude
\begin{equation}
\left\langle{\rm out}|{\rm in}\right\rangle\Big\vert_{J=0}
=\exp\left(i\Gamma/\hbar\right)\big\vert_{J=0}
\label{12}
\end{equation}
\end{enumerate}

Let us rewrite the definition of the effective action. To define
the path integral it is convenient to make a so-called Wick rotation, or
Euclidization, i.e. one replaces the real time coordinate to the purely
imaginary one $t \to -it$ and singles out the imaginary factor also from the
action $S \to -iS$ and the effective action $\Gamma \to -i\Gamma$. Then the
metric becomes Euclidean, i.e. positive definite,
and the classical action in all  good field theories becomes positive
definite functional.
So, the Euclidean effective action is defined to satisfy the equation
\begin{equation}
\exp\left({1\over \hbar}\Gamma(\phi)\right)=
\int\! {\rm D}\varphi\ \exp\left\{{1\over \hbar}
\left[S(\varphi)-(\varphi^i-\phi^i)\Gamma_{,i}(\phi)\right]\right\}
\label{13}
\end{equation}

This path integral is still not well defined. There is not any reasonable
method, except for lattice theories, to calculate this integral in general
case. The only path integrals that can be well defined are the Gaussian ones. Then 
the full path integral can be well defined as an asymptotic series of Gaussian
ones. This is just the quasiclassical, or WKB, approximation in the usual
quantum mechanics. We decompose the fields in the classical and quantum parts
\begin{equation}
\varphi=\phi+\sqrt\hbar h
\label{17}
\end{equation}
and look for a solution of the equation for the effective action in form of an
asymptotic series in powers of Planck constant.
\begin{equation}
\Gamma(\phi)=S(\phi)+\sum\limits_{n \ge 1}\hbar^n \Gamma_{(n)}(\phi)
\label{18}
\end{equation}Then all the coefficients of this expansion can be found. They are expressed in
terms of the well-known Feynman diagrams. The number of loops in these diagrams
correspond to the power of the Planck constant. We will be interested below in
the so called {\it one-loop effective action}
\begin{equation}
\Gamma_{(1)}=-{1\over 2}\ln{\rm Det}\, F
\label{19}
\end{equation}
where $F_{ik}=S_{,ik}$.

\section{Determinants of elliptic operators}

\setcounter{equation}{0}

Although Eqn.(\ref{19}) seems to be very easy it is still ill defined. The point
is it is divergent. This is just the well-known ultraviolet divergence of the
quantum field theory. Indeed, we can rewrite the functional determinant as
\begin{equation}
\ln\det A = {\rm Tr}\,\ln A =\ln \prod_n\limits \lambda_n 
=\sum\limits_n\ln\lambda_n
\label{20}
\end{equation}
where $\lambda_n$ are the eigenvalues of the operator $A$. This series is easy
to show to be divergent.

That means that one needs a regularization. This point was investigated very
thoroughly by many authors and it is found that in quantum gravity and gauge
theories the most appropriate regularizations are the analytical ones. The
functional determinants can be well defined in terms of the so called
$\zeta$-function.

At first let us use a  Wick rotation to produce
a Laplace operator from the wave operator. It allows to make use of methods for the 
evaluation of the determinants of elliptic operators \cite{Schwarz}. Now we are ready
to introduce the $\zeta$-function of an elliptic operator $A$:
$$
\zeta (s,A) = \sum_{i} \lambda_{i}^{-s}
$$
where $\{ \lambda_{i} \}$ are eigenvalues of the operator $A$. Then
$$
{\rm Tr}\,\ln A = - \left.\frac{\rm d}{{\rm d}s}\zeta(s,A)\right|_{s=0}
$$ 
To study the behavior of the $\zeta$-function it is common to use the formula
$$
 \lambda_{i}^{-s} =  \frac{1}{\Gamma(s)}\int\limits_0^\infty\! {\rm d}t\ 
t^{s-1} \exp(-t\lambda_{i}) 
\ .
$$
Summing over all nonzero eigenvalues of $A$ we get
\begin{equation}
\zeta (s,A) =  \frac{1}{\Gamma(s)}\int\limits_0^\infty\!{\rm d}t\ t^{s-1} 
( {\rm Tr}\, \exp(-tA) - L(A))
\label{b}
\end{equation}
where $L(A)$ is a number of zero-modes of $A$. 
Representation (\ref{b}) is valid when the integral converges.
For nonnegative operator it always does converge as $t \rightarrow +\infty$.
Conditions of the convergency of the integral as $t \rightarrow +0$ depend
on details of the operator $A$. For Laplace operator in 4D it converges as 
$t \rightarrow +0$ if ${\rm Re}\,s > 2$.

There exists the well-known Seeley expansion for the ${\rm Tr}\,(\exp(-tA))$:
$$
{\rm Tr}\,\exp(-tA) = \sum_{k\geq 0} \Phi_{-k}(A) t^{-k} + \rho(t) \ ,
$$
where $|\rho(t)|$ is bounded by a constant times $t$ as $t \rightarrow +0$.
$\Phi_{-k}(A)$ are called Seeley coefficients and play important role in the investigation of elliptic operators and their topological properties (for example in the Index theory). For example, for 4D Laplace operator $\Phi_{-k} = 0$
$k\geq 3$, and
$$
\begin{array}{l}
\Phi_{-2} = (4\pi)^{-2} \ , \\
\Phi_{-1} = - (4\pi)^{-2}\frac{1}{6} R \ , \ 
\mbox{ where $R$ --- scalar curvature}\ ,
\end{array}
$$
and
$$
\Phi_{0} = (4\pi)^{-2} \left( -\frac{1}{30}\nabla^2 R + \frac{1}{72}R^2 -
\frac{1}{180}R_{\mu\nu} R^{\mu\nu}
 + \frac{1}{180}R_{\mu\nu\sigma\rho} R^{\mu\nu\sigma\rho} \right) \ .
$$
Here $R_{\mu\nu\sigma\rho}$ and $R_{\mu\nu}$ 
are Riemann and Ricci tensors correspondingly. 
They are defined by a metric of a curved space. The latter coefficient $\Phi_0$ 
is particular important for the calculation of determinants.

Splitting (\ref{b}) into an integral over $[0,1]$ and one over $[1,+\infty]$ we get
\begin{eqnarray*}
\zeta(s) & = & \frac{1}{\Gamma(s)} \biggl( \sum_{k>0} 
\frac{\Phi_{-k}(A)}{s-k} +
\frac{\Phi_{0}(A)-L(A)}{s} + \\
& + & \int_1^{\infty}\!{\rm d}t\ {\rm Tr}\,\exp(-tA) t^{s-1}  + 
\int_0^{1}\!{\rm d}t\  \rho(t) t^{s-1} \biggr)
\end{eqnarray*}
The singularity at $s=0$ turns out to be removable since 
$\lim_{s\rightarrow 0} s\Gamma(s) = 1$ and $\zeta(s)$ is defined by analytical
continuation from the half-plane ${\rm Re}\, s > 0$.

From the relation $\ln\det A = -\zeta^{\prime}(0)$ we find that
\begin{eqnarray*}
\ln\det A & = &  \sum_{k>0} \frac{\Phi_{-k}(A)}{k} +
\Gamma^{\prime}(1)\Bigl( \Phi_{0}(A) - L(A) \Bigr) - 
 \int_1^{\infty}\!\frac{{\rm d}t}{t} {\rm Tr}\,\exp(-tA)  - \\  
 & - & \int_0^1\!\frac{{\rm d}t}{t} \left( {\rm Tr}\,\exp(-tA) 
- \sum_{k<0} \Phi_{-k}(A) t^{-k}
\right) \ .
\end{eqnarray*}
This is $\zeta$-regularized $\ln\det A$ which we use in the paper.

To conclude this section we give the relation between $\ln\det A$ and
$\ln\det \alpha A$ where $\alpha$ is a number parameter. It is easy to see that
$$
\zeta(s,\alpha A) = \zeta(s,A) \cdot \alpha^{-s}
$$
because $\lambda_i(\alpha A) = \alpha \lambda_i(A)$. This leads to the relation:
\begin{equation}
\ln\det \alpha A = \ln\det A + \ln\alpha\cdot\zeta(0,A) =
\ln\det A + \ln \alpha\cdot\Bigl(\Phi_0(A) - L(A) \Bigr)
\end{equation}
which we use intensively in the paper.

\section{Variational derivative of the effective action}
\setcounter{equation}{0}

In this appendix we give details of the calculation omitted in section 3
to do not interrupt the main stream of the consideration. To make the calculation
more efficient we make use covariant form of the quantum correction since
there exists well-know way to simplify covariant variation calculation
\cite{DNFW}.

We are interested in the following variational derivative
\begin{equation}
\frac{\delta }{\delta g^{\mu\nu}} \Phi_{0} (\Box )
\end{equation}
where
\begin{equation}
\Phi_{0} (\Box ) =
\frac{1}{(4\pi)^2} \int \!{\rm d}x\ \sqrt{-g}
\left(\frac{1}{72}R^2 -
\frac{1}{180}R_{\mu\nu} R^{\mu\nu}
 + \frac{1}{180}R_{\mu\nu\sigma\rho} R^{\mu\nu\sigma\rho} \right) 
\end{equation}
For variation of the volume element we have
$$
\delta\sqrt{-g} = - \frac{1}{2\sqrt{-g}}\delta g_{\mu\nu}\Delta^{\mu\nu} =
\frac{\sqrt{-g}}{2}\delta g_{\mu\nu}g^{\mu\nu}
 = -\frac{\sqrt{-g}}{2}g_{\mu\nu}\delta g^{\mu\nu}
$$
where $\Delta^{\mu\nu}$ is defined by as
$$
\Delta^{\mu\nu} = g\cdot g^{\mu\nu}
$$
and the
following relation was used
$$
\delta g_{\mu\nu}g^{\mu\nu} = -g_{\mu\nu}\delta g^{\mu\nu}\ .
$$
For variations 
$R^2,R_{\mu\nu} R^{\mu\nu},R_{\mu\nu\sigma\rho} R^{\mu\nu\sigma\rho}$ 
one can obtain
\begin{eqnarray*}
{\delta} R^2 \!\!&=&\!\! 2\delta R\cdot R = 2\delta g^{12}R_{12} R 
+ 2 g^{12}\delta R_{12}\cdot R
\\
{\delta} R_{12} R^{12} \!\!&=&\!\! \delta (g^{13}g^{24}R_{12}R_{34}) =
2\delta g^{12} R_{13} R_{2}^{\,\cdot\, 3} + 2 \delta R_{12} R^{12}
\\
{\delta} R_{1234} R^{1234} \!\!&=&\!\! \delta (g^{15}g^{26}g^{37}g^{48}R_{1234}R_{5678}) =
4\delta g^{12} R_{1345} R_{2}^{\,\cdot\, 345} + 2 \delta R_{1234} R^{1234}
\end{eqnarray*}
where numbers $1,2,3,\dots$ mean indices $\mu_1,\mu_2,\mu_3,\dots$ respectively.
Hence,
$$
\delta \Phi_0 = \frac{1}{36(4\pi)^2}\int\!{\rm d}x\ \sqrt{-g}\,\delta g^{\mu\nu} 
\Biggl[
- \frac{1}{2}g_{\mu\nu}
\Biggl\{\frac{1}{2}R^2 - \frac{1}{5}R_{12}R^{12} + \frac{1}{5}
R_{1234}R^{1234}\Biggr\} 
$$
$$
+ R_{\mu\nu}R
 - \frac{2}{5}R_{\mu 1}R_k^{\,\cdot\,\nu}
 + \frac{4}{5}R_{\mu 123}R_\nu^{\,\cdot\, 123}
\Biggr]
$$
$$
+ \frac{1}{36(4\pi)^2}\int\!{\rm d}x\ \sqrt{-g} \Biggl[
g^{12}\delta R_{12} R - \frac{2}{5} \delta R_{12} R^{12} + \frac{2}{5}
\delta R_{1234} R^{1234} \Biggr]
$$
So we need to calculate last three terms. 

For variations of Ricci and Riemann tensors we have \cite{DNFW}:
\begin{eqnarray}
\delta R_{\mu\nu} &=& 
   \nabla_\nu(\delta \Gamma^1_{\mu 1})
 - \nabla_1(\delta \Gamma^1_{\mu\nu})
\nonumber\\ 
&=&
\frac{1}{2}g^{12}\left[ 
   \nabla_\mu\nabla_\nu\delta g_{12}
 - \nabla_1\nabla_\mu\delta g_{2\nu}
 - \nabla_1\nabla_\nu\delta g_{2\mu}
 + \nabla_1\nabla_2\delta g_{\mu\nu}
\right]
\label{varricci}\\
\delta R_{\mu\nu\sigma\rho} &=& 
 - g_{\mu1}\delta g^{12} R_{2\nu\sigma\rho}
 + g_{\mu1}\nabla_\rho(\delta \Gamma^1_{\nu\sigma})
 - g_{\mu1}\nabla_\sigma(\delta \Gamma^1_{\nu\rho})
\nonumber\\ 
&=&
 - g_{\mu1}\delta g^{12} R_{2\nu\sigma\rho}
 + \frac{1}{2}
\Bigl[
   \nabla_\rho\nabla_\sigma\delta g_{\mu\nu}
 + \nabla_\rho\nabla_\nu\delta g_{\mu\sigma}
 - \nabla_\rho\nabla_\mu\delta g_{\nu\sigma}
\nonumber\\ 
&&
 - \nabla_\sigma\nabla_\rho\delta g_{\mu\nu}
 - \nabla_\sigma\nabla_\nu\delta g_{\mu\rho}
 + \nabla_\sigma\nabla_\mu\delta g_{\nu\rho}
\Bigr]
\label{varriemann}
\end{eqnarray}
Let us consider covariant divergence of some vector $T^k$:
$$
\nabla_\mu T^\mu = \nabla^\mu T_\mu = \partial_\mu T^\mu 
+\Gamma_{\mu\nu}^\nu T^\mu = 
\frac{1}{\sqrt{-g}}\partial_\mu(\sqrt{-g}T^\mu)
$$
Hence
$$
\sqrt{-g}\nabla_\mu T^\mu = \partial_\mu T^\mu
$$
is a pure divergence and can be dropped under integration.
Using this we can easily derive the following expression for variations
of the integrals we are interested in
\begin{eqnarray*}
&\!\!\!&\!\!\!\!\!\!
\int\!{\rm d}x\ \sqrt{-g} g^{\mu\nu}\delta R_{\mu\nu} R =
\int\!{\rm d}x\ \sqrt{-g} \delta g^{\mu\nu}
\left[ 
 - \frac{1}{2}\{\nabla_\mu,\nabla_\nu\}R
 + g_{\mu\nu}\Box R
\right]\\
&\!\!\!&\!\!\!\!\!\!
\int\!{\rm d}x\ \sqrt{-g} \delta R_{\mu\nu} R^{\mu\nu} = 
\frac{1}{2}\int\!{\rm d}x\ \sqrt{-g} \delta g^{\mu\nu}
\left[
 - \Box R_{\mu\nu}
 - g_{\mu\nu}\nabla_1\nabla_2 R^{12}
 + \nabla_1\nabla_\mu R_\nu^{\,\cdot\,1}
 + \nabla_1\nabla_\nu R_\mu^{\,\cdot\,1}
\right]\\
&\!\!\!&\!\!\!\!\!\!
\int\!{\rm d}x\ \sqrt{-g} \delta R_{\mu\nu\sigma\rho} R^{\mu\nu\sigma\rho} = 
\int\!{\rm d}x\ \sqrt{-g} \delta g^{\mu\nu}
\left[
 - R_{\mu123} R_\nu^{\,\cdot\,123}
 + 2\nabla^2\nabla^1 R_{\mu12\nu}
\right]
\end{eqnarray*}
Summing all terms we obtain
\begin{eqnarray*}
\frac{\delta }{\delta g^{\mu\nu}} \Phi_{0} (\Box )\!\!& =\!\! &
 \frac{1}{36(4\pi)^2}
\sqrt{-g} 
\Biggl\{
 - \frac{1}{2} g_{\mu\nu} 
\Bigl[ 
   \frac{1}{2}R^2
 - \frac{1}{5} R_{\sigma\rho}R^{\sigma\rho} 
 + \frac{1}{5}R_{\sigma\rho\alpha\beta}R^{\sigma\rho\alpha\beta}
\Bigr] 
\\
\!\!& +\!\! & 
   R_{\mu\nu}R
 - \frac{2}{5}R_{\mu\sigma}R_{\nu}^{\,\cdot\,\sigma} 
 + \frac{2}{5}R_{\mu\sigma\rho\alpha}R_{\nu}^{\,\cdot\,\sigma\rho\alpha}
 - g_{\mu\nu} \Box R
 + \frac{1}{2} \{\nabla _{\mu},\nabla_{\nu}\} R  
\\
\!\!& +\!\! &
\frac{1}{5} 
\Bigl[ 
 - 2\nabla^{\sigma}\nabla_{\nu} R_{\mu\sigma}
 + \Box R_{\mu\nu}
 + g_{\mu\nu}\nabla^{\sigma}\nabla^{\rho} R_{\sigma\rho} 
\Bigr]
 + \frac{4}{5} 
   \nabla^{\rho}\nabla^{\sigma}R_{\mu\sigma\rho\nu} 
\Biggr\}
\end{eqnarray*}


\newpage

\begin{center}
Figure caption.
\end{center}

\begin{enumerate}
\item Fig.1 \ Bose condensate density profile obtained in mean field
approximation. The curve is drawn using the analytic Ansatz of Ref.\cite{IM}.
Lengths are measured in units of the oscillator length $a_{\perp}$. 
The multiplier is defined as $\lambda = 4\pi l/a_{\perp}$.
\item Fig.2 \ The calculated  one-loop quantum correction to
the Nonlinear Schr\"odinegr equation in the central region of the 
isotropic trap. Lengths are measured in units of the oscillator length.
$N$ is the number of particles (for the picture $N=10^5$) and
$\delta P$ is defined by eq.(\ref{delta}).
\item Fig.3 \ The calculated  one-loop quantum correction to
the Nonlinear Schr\"odinegr equation at the intermediate region of the 
isotropic trap.  Lengths are measured in units of the oscillator length.
$N$ is the number of particles (for the picture $N=10^5$) and
$\delta P$ is defined by eq.(\ref{delta}).
\item Fig.4 \  Change in sign of the one-loop contribution
in intermediate region of the trap.  Lengths are measured in units of the oscillator length. $N$ is the number of particles (for the picture $N=10^5$) 
and $\delta P$ is defined by eq.(\ref{delta}).
\item Fig.5 \ Merging of High density approximation for the quantum correction
(solid line) and the corresponding Low density one (dashed line). 
Lengths are measured in units of the oscillator length.
In the region to the left to the point $r\sim 6.13$ High density
approximation produces the correction (\ref{delta}).
In the region to the right to the  point $r\sim 6.13$ Low density
approximation (\ref{pl}) for the correction should be used.
In the region of the boundary point both approximations fail and 
other quantum contributions should be kept.
\item Fig.6  \  Comparison of various terms in the equation for the
condensate density. The interpolating quantum correction (solid line) of Fig.5, nonlinear interaction term (dashed line) and kinetic energy (circles) are pictured. 
Lengths are measured in units of the oscillator length.

\end{enumerate}

\begin{figure}
\centerline{\epsfxsize=16cm \epsfbox{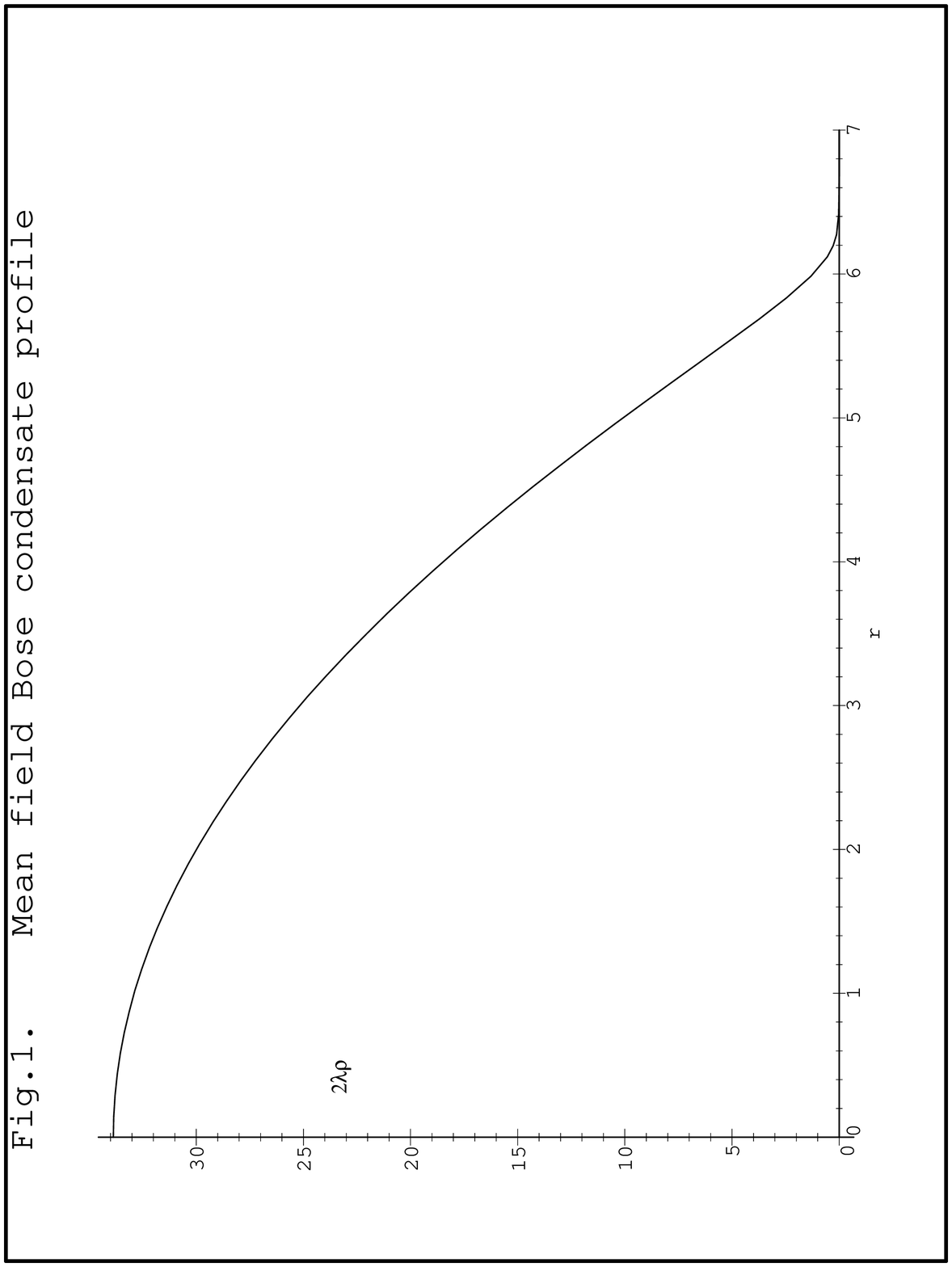}}
\label{kir1}
\end{figure}

\begin{figure}
\centerline{\epsfxsize=16cm \epsfbox{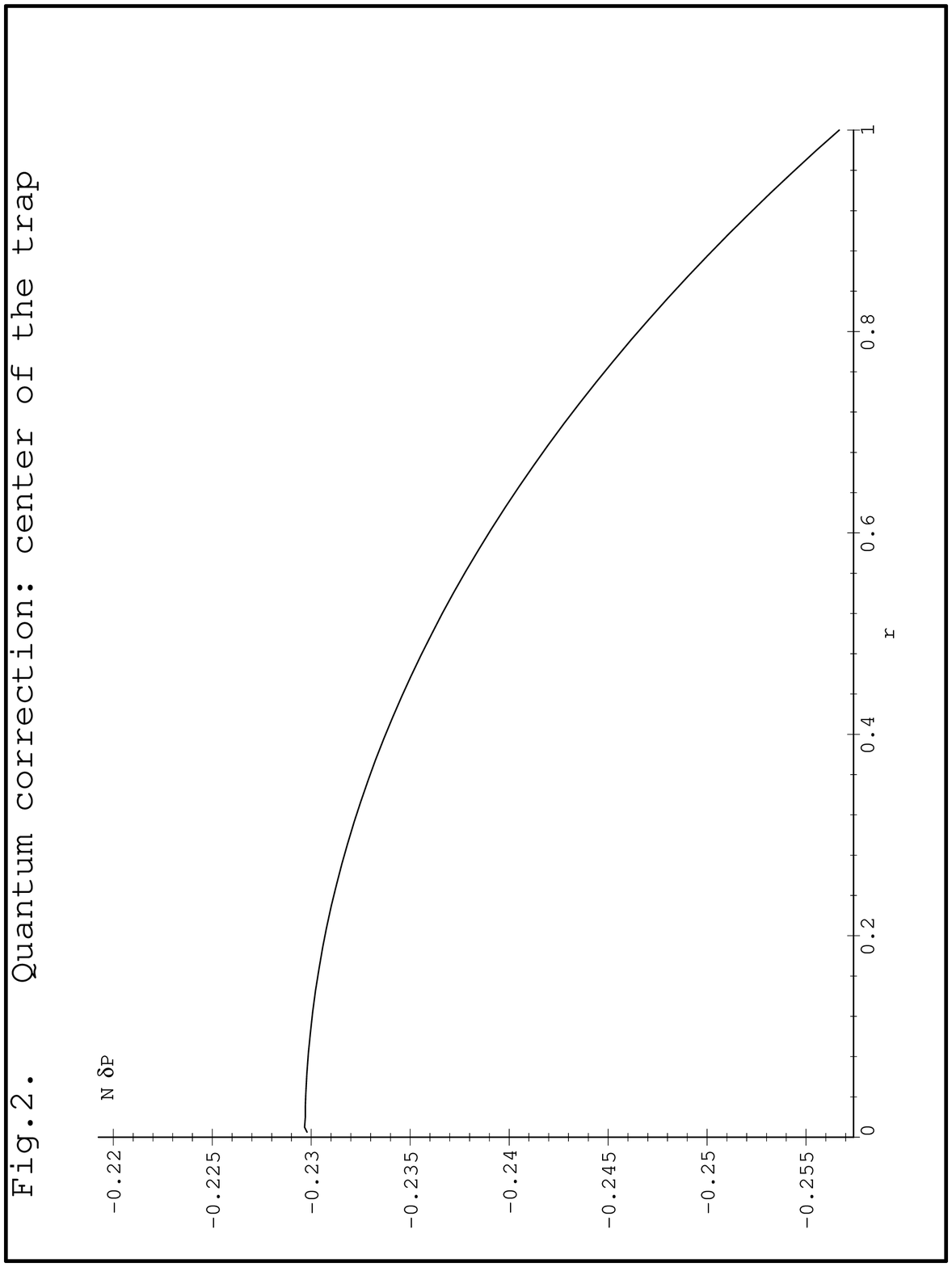}}
\label{kir2}
\end{figure}

\begin{figure}
\centerline{\epsfxsize=16cm\epsfbox{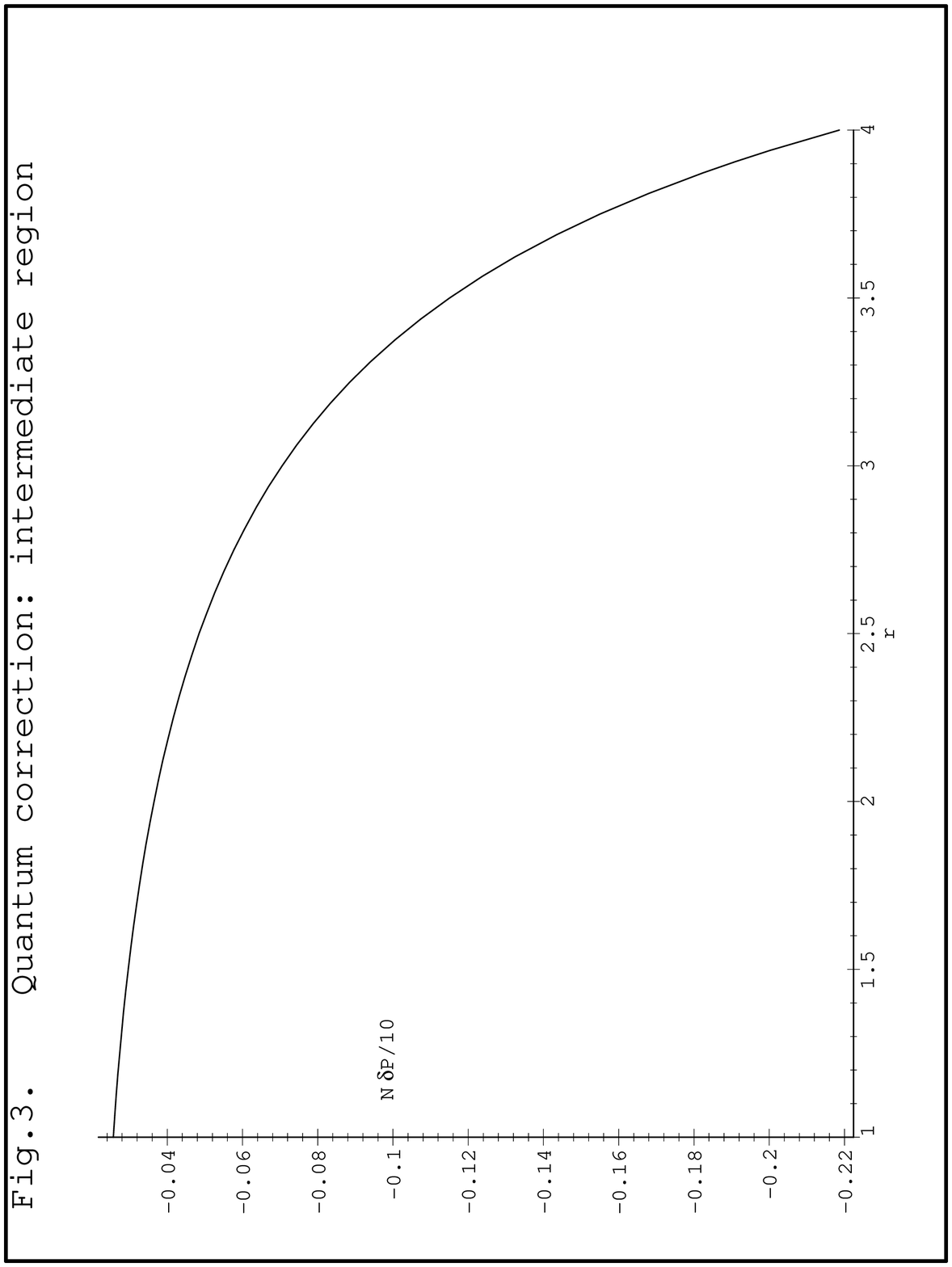}}
\label{kir3}
\end{figure}

\begin{figure}
\centerline{\epsfxsize=16cm \epsfbox{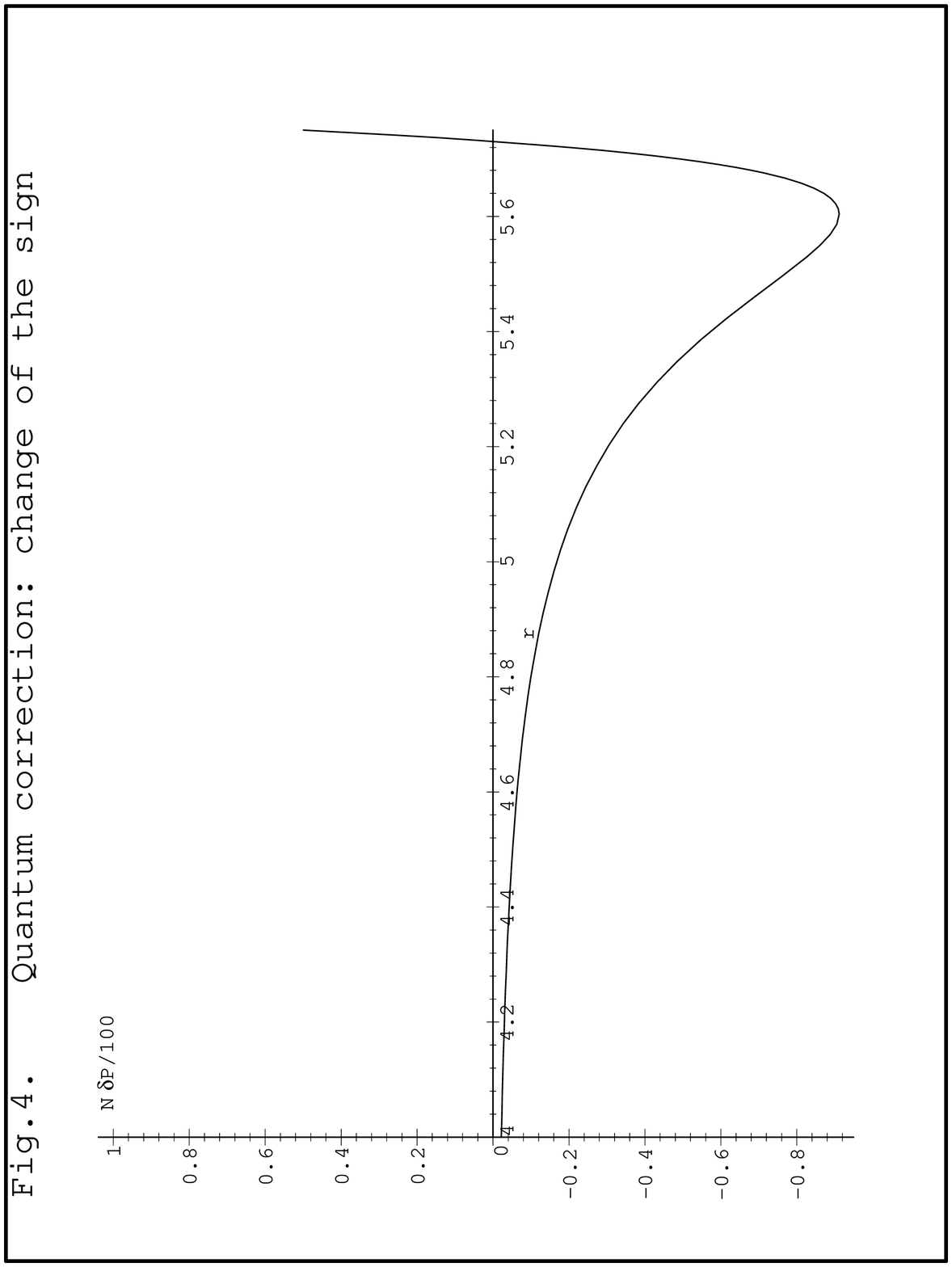}}
\label{kir4}
\end{figure}

\begin{figure}
\centerline{\epsfxsize=16cm \epsfbox{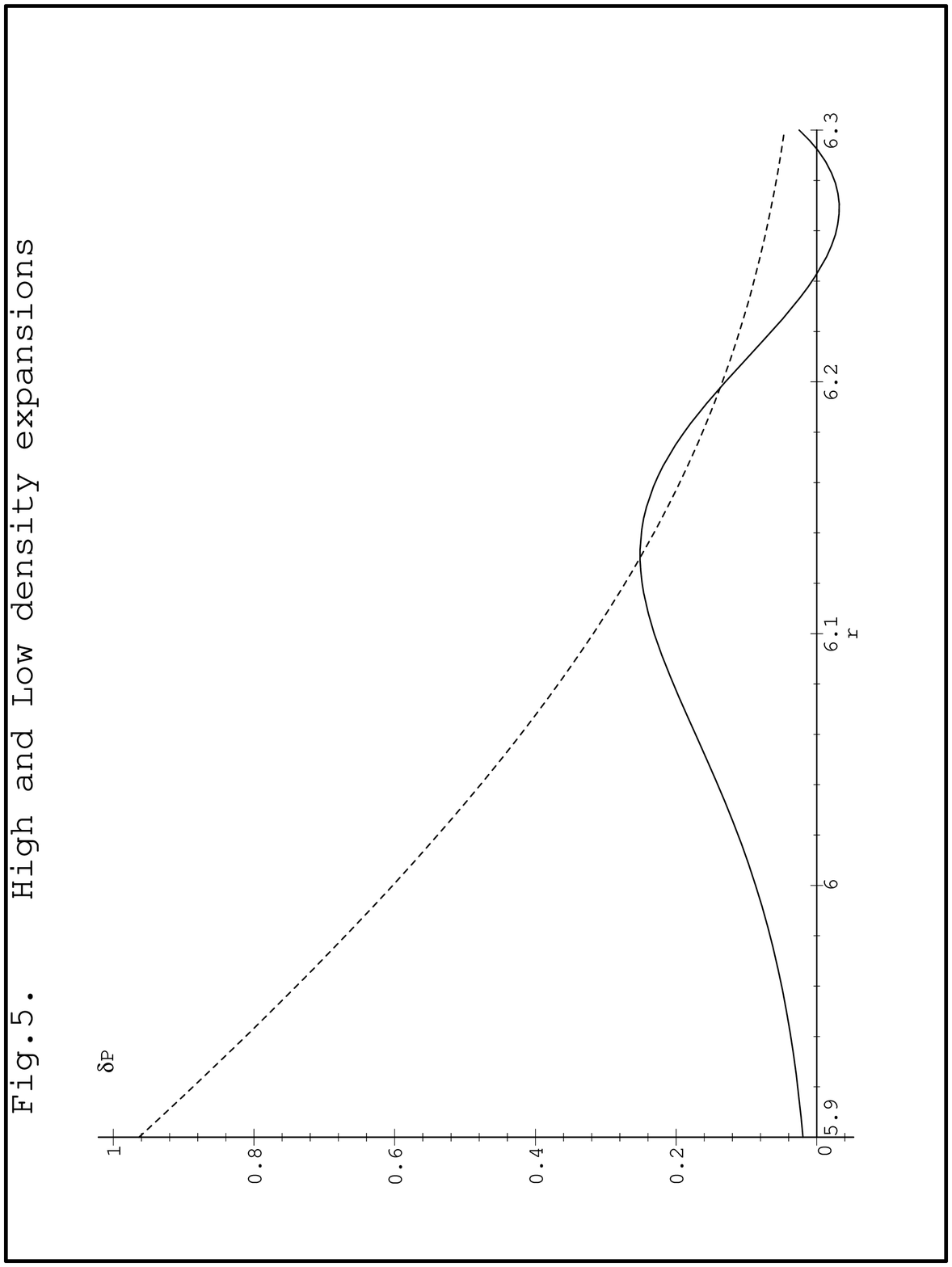}}
\label{kir5}
\end{figure}

\begin{figure}
\centerline{\epsfxsize=16cm\epsfbox{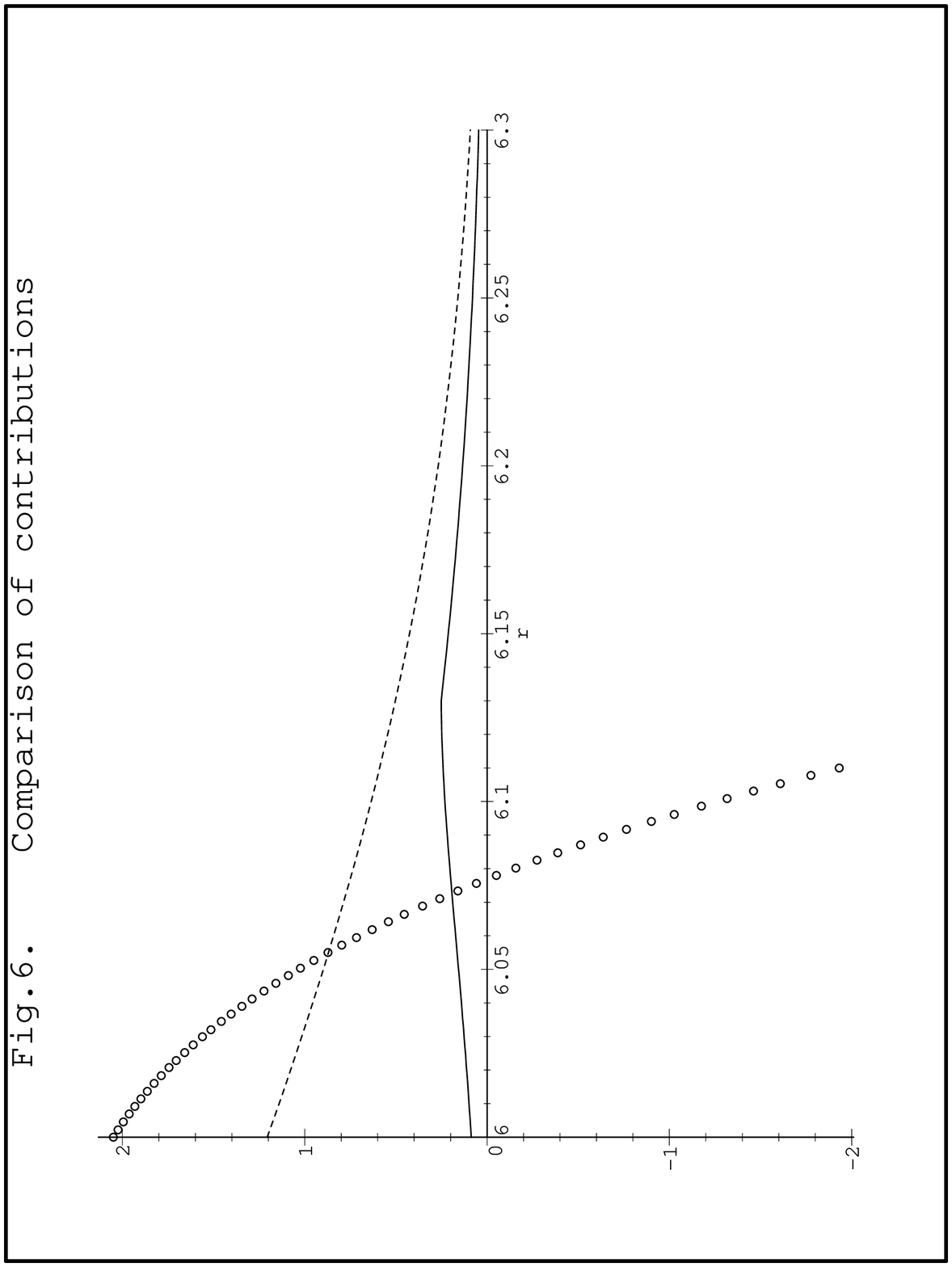}}
\label{kir6}
\end{figure}


\begin{thebibliography}{99}
\bibitem{note5} 
Popov approximation is applicable to incorporate the interaction of the condensate with the condensate if the condensate density is small.
This approach was used recently to investigate the thermodynamics 
of the 
trapped Bose gas near phase transition point \cite{Pit}.
\bibitem{Osc}The $s$-wave scattering length can be estimated as
$l \sim 100 a_{0}$, where
$a_{0}$ is the Bohr radius (i.e., $l \sim 5 \cdot 10^{-7} \mbox{cm}$);
\bibitem{note} Although in the section the isotropic case  is only considered
it is easy and straightforward to give the corresponding consideration for
anisotropic case. To do this the only change is to make use the analytic
expression for anisotropic condensate profile which can be found in \cite{IM}.
\bibitem{note3}
It is not absolutely true
for two-point correlation function only when the approach gives an partial derivatives differential equation which still has to be solved; it can be done
numerically and is not principal problem.
\end{thebibliography}
\end{document}